\documentstyle[aps,preprint]{revtex}
\draft

\begin{document}

\title{Pattern selection of cracks in directionally drying fracture}

\author{Teruhisa S. Komatsu}
\address{Department of Physics, Tohoku University, Sendai 980-77, Japan}

\author{Shin-ichi Sasa
\thanks{Present address:
Department of Physics, University of Illinois at Urbana-Champaign,
1110 West Green Street, Urbana, IL, 61801, USA
}
}
\address{
Department of Pure and Applied Sciences,
         College of Arts and Sciences, University of Tokyo,
         Komaba, Meguro-ku, Tokyo 153, Japan
}

\date{\today}
\maketitle

\begin{abstract}
We study pattern selection of cracks in
directionally drying fractures by analyzing  the experimental
systems recently devised by C. Allain and L. Limat.
[Phys. Rev. Lett. {\bf 74}, 2981 (1995).]
Proposing a simple picture of crack formation,
we clarify the mechanism of how cracks array regularly
and find that the interval between neighboring cracks is proportional to
the $2/3$ power of the cell thickness.
This result  explains well the experimental data
of Allain and Limat.
\end{abstract}





The study of fracture has developed greatly \cite{herrman}
since Griffith wrote a breakthrough paper in 1920 \cite{griffith}.
In particular, recent progress is due largely to the development of
well-controlled experimental systems \cite{fineberg,yusesano:vib}.
Recently, Allain and Limat have studied periodically aligned crack patterns
by devising an experimental system consisting of directionally drying fractures
\cite{allain}. In their experiment, a colloidal suspension was
put into a rectangular cell in which one surface was left open
in order to allow evaporation, and after a short time,
periodically aligned cracks were observed \cite{allain}.
Similar phenomena have been observed
in experiments on drying fracture
with different geometries \cite{nakahara}.
Since a crack cannot adjust its position
after it appears, these experimental results
cannot be explained in the same way
as periodic pattern formations in convective systems
and reaction diffusion systems \cite{crosshohenberg}.
We are thus led to consider the mechanism of crack  formations
in drying fracture.


The question we address is to determine the interval between
neighboring cracks, $\lambda$.
Although Allain and Limat gave a theoretical estimate of $\lambda$ and
confirmed that its value is of the same order as the experimental
result \cite{allain}, their theory leads to a relation between $\lambda$
and the system thickness $H$  as $\lambda/H \sim A-B\ln H $,
which does not fit their experimental data well.
In this Letter, we propose a simple picture of
crack formation in directionally drying fracture. Based on this picture,
we clarify the reason why cracks are formed
periodically and derive the scaling relation  $\lambda \sim H^{2/3}$.
Further, by comparison with the experimental data,
we confirm the validity of the scaling relation.


The experimental configuration we analyze is illustrated in Fig.1.
We assume that the cell under consideration extends semi-infinitely in
the positive $y$ direction and has boundaries at $y=0$ (front surface),
$x = \pm  L/2$, and $z= \pm H/2$. Further, we are interested in the
limiting case  $H \ll L$.


We first consider the water distribution in the material
without cracks. Let $\phi$ be the water volume fraction. We assume that
water evaporates from the front surface at a rate $(\phi-\phi_\infty)J$,
where $\phi_\infty$ is the equilibrium value of the water volume fraction,
and that there is no flux at the other boundaries.
In a bulk region, a diffusion current proportional
to the gradient of $\phi$ is assumed to arise. Then, the time evolution
of $\phi$ is given by the diffusion equation
\begin{equation}
\frac{\partial \phi}{\partial t}=D \triangle \phi,
\label{eqn:diff}
\end{equation}
where $D$ is a diffusion constant. As an initial condition,
we assume that $\phi$ takes a constant value $\phi_0 > \phi_\infty$.
Then, $\phi$ is independent of $(x,z)$, and  $\phi(y,t)$ is expressed by
using a Green function of the diffusion equation.
Without knowing the explicit form of $\phi(y,t)$,
we do know the following general aspects of its
behavior which are important for the argument given below.
First, when $t$ is fixed and $y$ is increased,
$\phi(y,t)$ monotonically approaches $\phi_0$
with the characteristic variation scale $\sqrt{Dt}$.
Second,
$\phi(0,t)$ approaches $\phi_\infty$ with the time scale $t_e=D/J^2$.
Then, the length scale reached by the diffusion due to the evaporation
at the front surface, which is denoted by $\xi$, is estimated
as $\xi=\sqrt{D t_e}=D/J$.
In the argument below, $\xi \gg  L$ will be  assumed
so that we can concentrate on the idealized case
that  the evaporation at crack surfaces does not
cause inhomogeneity in the $x$ direction.


We next discuss  elastic properties of the material.
{}From a macroscopic viewpoint, the material can be regarded as
a homogeneous elastic medium.  We thus apply the linear elastic theory
to the calculation of  the macroscopic stresses.
Here, the stresses $\sigma_{xx}$, $\sigma_{yy}$ and
$\sigma_{zz}$  are proportional to the sum of the corresponding strains
and the volume shrinkage rate, $C$,  due to the evaporation
under the stress free boundary condition, where $C$ is assumed to be given by
\begin{equation}\label{cphi}
 C  = - \alpha \frac{ \phi- \phi_0 }{\phi_0}.
\end{equation}
We  note that elastic constants and $\alpha$ may depend on
the volume fraction of suspensions, which is fixed at $t=0$, but
is not strongly dependent on $\phi$.
Since  elastic fields vary  much faster than the diffusion
field $C$, values of stress fields are determined
adiabatically by the profile $\phi$. Then, in principle, we can calculate
stresses for a given $\phi(y,t)$ under given boundary conditions.


Now, we consider  a ``macroscopic'' crack formation.
In the system under consideration,
the seed of a crack is not supplied externally, but rather there are
microscopic inhomogeneities at the length scale of the particle radius.
One may regard the inhomogeneity  as micro-cracks embedded randomly in
the elastic medium. If we assume such a picture, the Griffith criteria
can be applied to the understanding of macroscopic
crack formation \cite{griffith}.
That is, a micro crack can grow if the energy release rate (per unit length)
for the micro crack growth exceeds the surface energy (per unit length),
and when the energy release rate increases
further as the micro-crack grows, the crack grows acceleratedly
and finally becomes macroscopically observable.
This process corresponds to
a macroscopic crack formation.
However, due to the randomness of
the positions of micro cracks, it is difficult to calculate precisely
the energy release rate of each crack. Thus, a coarse-grained picture is
needed. Since we believe that macroscopic crack formation does not depend on
microscopic details and can be described by physical quantities
defined on the macroscopic scale, we propose a following hypothesis:
The relevant quantity for macroscopic crack formation is the
sum of the elastic energy of macroscopic stress fields and the surface energy,
and if the total energy decreases with crack formation, the crack appears
macroscopically.  Inspection of  the  hypothesis from a microscopic viewpoint
may be an important study, but here we proceed to a discussion
of crack formation processes based on this hypothesis.


We discuss the time evolution of the system.
First, the material tends to shrink due to the evaporation.
However, since the displacement at the horizontal boundaries is zero,
internal stress proportional to  $C$  are created.
In the case that the material is sufficiently thin,
we expect that a crack is formed along the $z$-direction at the front surface
where the stress takes a maximum value.
The crack breaks the front surface and extends along the $y$ direction
until the energy release rate for the crack extension is
equal to the surface energy.  Therefore, the crack spacing observed
in the experiment \cite{allain} is determined during the crack formation
process at  the front surface. Hereafter, we will discuss the crack formation
at the strip corresponding to the front surface.
Further, for simplicity,
we assume that separations at the vertical boundaries ($x=\pm L/2$)
occur before crack formations in the bulk.
Note that the system after such separation is equivalent to
one under stress free conditions at the vertical boundaries.
Whether this assumption is realistic or not
depends on the conditions at the interfaces
between the elastic material and the cell.
However, as we will see later,
even in the case that the separation from the boundaries never occurs,
the result of pattern selection of cracks is unchanged.

As the evaporation proceeds, crack formation occurs first
at a time $t_1$ satisfying the equation
\begin{equation}
\max_{l}[ E(L,C(t_1))-E(l, C(t_1))-E(L-l, C(t_1))] = \Gamma H,
\label{eqn:con1}
\end{equation}
where $\Gamma$ is the surface energy per unit length, and
$E(L,C(t))$ is the elastic energy of the material with the horizontal
length $L$ and the volume shrinkage rate $C(t)$ at time $t$.
The value of  $l$ maximizing the quantity
$E(L, C(t_1))-E(l, C(t_1))-E(L-l, C(t_1))$, denoted by $L_1$,
specifies the position where the crack is formed.
{}From the geometrical symmetry of the problem, we expect $L_1=L/2$.
This implies that a crack is formed first at the center of the strip.
(This will be confirmed later.)

In this way, at a time $t_1$,  there are two stripes with a horizontal
length $L/2$. We should notice here that each strip has the same boundary
conditions as the original one. Thus, by replacing variables $(t_1, L)$ in
Eq.(\ref{eqn:con1}) with $(t_2, L_1)$, we know the
time $t_2$ at which the next crack formations occur at the center of each strip
with the length $L_1/2$.
(Note that these crack formations occur simultaneously.)
As bisected strips are further bisected in succession,
this process repeats, producing strips with equal horizontal length
until the evaporation finishes.
This is the reason the cracks form periodically in space.
The interval between cracks $\lambda$,
which is identical to the horizontal size of strips at $t=\infty$,
is determined by the maximum length $ 2^{-n} L \; (n:\mbox{integer})$
shorter than the length $\lambda_{*}$ satisfying
\begin{equation}
 E(2\lambda_{*}, C_\infty)-2E(\lambda_{*}, C_\infty) = \Gamma H.
\label{eqn:lambdas}
\end{equation}
Also, as easily checked, $\lambda$ satisfies the inequality
\begin{equation} \label{lambdavslambdastar}
\frac{\lambda_*}{2} < \lambda < \lambda_*.
\label{eqn:lambda}
\end{equation}
Therefore, we can estimate the value of $\lambda$ if
we can succeed in deriving an expression for the elastic energy of
the strip. In the following paragraphs,
we will derive an expression for $E(l, C)$.


In order to evaluate the expression of the elastic energy of the strip
with a horizontal size $l$, we propose to consider a quasi-one dimensional
spring network which is composed of a chain of $N$ springs
along the center line and vertical springs connecting each node to a fixed
position at the boundary (see Fig.\ref{graph:effecone}).
The ends of the horizontal springs have no constraints
because free boundary conditions are imposed at the vertical
boundaries of the strip.
Here, the natural lengths of horizontal and vertical springs at $t=0$
are given by $a=l/N$ and $H/2$, respectively, and
the attachment points are assumed to be positioned regularly, with
a period $a$. Further, the spring constants of horizontal and vertical springs
are denoted by $k_1$ and $k_2$ respectively.
This effective spring network resembles Meakin's model \cite{meakin}
when the model is supplemented with a breaking rule.
We note, however, that
according to our picture of crack formation discussed above,
the horizontal spring is broken deterministically,
as in Hayakawa's model
\cite{hayakawa:vib}, not probabilistically, as in Meakin's model.

The elastic energy of the spring network is expressed by
\begin{equation}\label{energy:start}
\sum_{i=0}^{N-1} \frac{k_1}{2}
\left[ u_{i+1}-u_{i} - a \frac{C}{3} ) \right]^2
+2 \sum_{i=0}^{N} \frac{k_2}{2}
\left[ \sqrt{ u_i^2 + \left(\frac{H}{2}\right)^2}
	 - \frac{H}{2} (1-\frac{C}{3}) \right]^2.
\end{equation}
Here, $u_i$ is the displacement of the $i$-th node from
the reference point $ x_i \equiv ia - l/2$.
Note that the linear shrinkage rate of springs is given by $C/3$.
Under the assumption $ (u_i/H)^2 \ll C/3 \ll 1 $,
expanding  Eq.(\ref{energy:start}) in $u_i/H$ and
ignoring terms of higher order than $(u_i/H)^2$,
Eq.(\ref{energy:start}) reduces to
\begin{equation}\label{energy:discrete}
\sum_{i=0}^{N-1} \frac{k_1}{2}
\left[ u_{i+1}-u_{i} + a \frac{C}{3}\right]^2
+\sum_{i=0}^{N} k_2 \left[ \left(\frac{CH}{6}\right)^2 + \frac{C}{3} u_i^2
\right].
\end{equation}
Further, in order to make it possible to develop an analytical argument,
we take the continuum limit ($a \rightarrow 0$ with fixing $L$)
of Eq.(\ref{energy:discrete}).
First, we introduce a variable $\tilde u$ well-defined in this limit by
\begin{equation}
u_i = H \tilde{u}\left(\frac{x_i}{H}\right).
\end{equation}
Then, when this expression is substituted into Eq.(\ref{energy:discrete}),
$E$ should not depend on $a$  in this limit.  By noting that
$u_{i+1}-u_i=a\tilde{u}'+O(a^2)$
where the prime refers to differentiation with respect to the argument $x_i/H$,
this requirement leads to the conditions $k_1 \sim 1/a$ and $k_2 \sim a$.
Further, since it seems natural to assume that
$k_1$ and $k_2$ do not depend on the horizontal length $l$,
we can express $k_1$ and $k_2$  in this limit as
\begin{equation}
k_1 =\frac{H}{a}K,\quad
k_2 = \frac{k_1}{2}\left(\frac{a}{H}\right)^2 \kappa,
\end{equation}
where $K$ is related to the Young modulus of the two dimensional elastic
material, and $\kappa$ is a non-dimensional quantity of order unity.
As a result, we obtain the following expression of the elastic energy
in the continuum limit:
\begin{equation}\label{energy:con}
\frac{KH^2}{2}\int^{l/2H}_{-l/2H}d\tilde{x}
\left[
\left(\tilde{u}'+\frac{C}{3}\right)^2+\frac{\kappa C}{3}\tilde{u}^2
+\frac{\kappa}{4}\left(\frac{C}{3}\right)^2
\right].
\end{equation}
Here, $\tilde{x}$ refers to  $x_i/H$ in the continuum limit.
The equilibrium value of the displacement field $\tilde u$ is obtained by
minimizing $E$. Thus, $\tilde u$ is given by a solution of the equation
\begin{equation}\label{dispeq:bulk}
\tilde{u}'' - \frac{\kappa C}{3} \tilde{u} =  0,
\end{equation}
under stress free boundary conditions at both side ends
$(\tilde{x}=\pm l/2H)$ :
\begin{equation}\label{dispeq:bc}
\tilde{u}' + \frac{C}{3} = 0.
\end{equation}
Solving this differential equation, we get an analytic expression for
the displacement field $\tilde u$ as
\begin{equation}\label{adisp}
\tilde{u}=-\sqrt{\frac{C}{3 \kappa}}
    \frac{ \sinh(\sqrt{\frac{\kappa C}{3}} \tilde{x}) }
         { \cosh(\sqrt{\frac{\kappa C}{3}} \frac{l}{2H}) }
,
\end{equation}
where $-l/2H\le\tilde{x}\le l/2H$.
Here, note that Eq.(\ref{adisp}) is valid only in the case
\begin{equation}\label{linearcon}
\sqrt{\frac{\kappa C}{3}}\frac{l}{H} \ll 1,
\end{equation}
because we have assumed $\tilde{u}^2\ll C/3 \ll 1$
in the derivation of Eq.(\ref{energy:discrete}).
By substituting Eq.(\ref{adisp}) into Eq.(\ref{energy:con}),
an expression for the elastic energy $E(l,C)$ is derived as
\begin{equation}\label{exact:energy}
E(l,C)=\frac{KC^2 H l}{18}
\left[\left(1+\frac{\kappa}{4}\right)
-\sqrt{\frac{3}{\kappa C}}\frac{2H}{l}
\tanh\left(\sqrt{\frac{\kappa C}{3}}\frac{l}{2H}\right)
\right].
\end{equation}
Further, under the condition Eq.(\ref{linearcon}),
Eq.(\ref{exact:energy}) becomes
\begin{equation}\label{energy:limit}
E(l,C)=\frac{K C^2 H l}{72}
\left[\kappa
+\frac{1}{9}\left(\sqrt{\kappa C}\frac{l}{H}\right)^2
\right].
\end{equation}
As expected, it is easily confirmed that
$E(L, C)-E(l, C)-E(L-l, C)$  is maximum when $l=L/2$,
because we find, using Eq.(\ref{exact:energy}), that
$\partial E(l,C)/ \partial l$ is a  monotonically increasing
function of $l$.
Thus from Eqs.(\ref{eqn:lambdas})and (\ref{energy:limit}), we obtain
\begin{equation}\label{lambdastar}
\lambda_*^3 = \gamma H^2,
\end{equation}
with
\begin{equation}
\gamma= \frac{108\Gamma}{K \kappa C_\infty^3}.
\end{equation}
Here, this relation is valid only when $ \sqrt{\kappa C/3}\lambda_*/H\ll 1$.
Using Eq.(\ref{lambdavslambdastar}),
the length $\lambda$ between cracks periodically aligned
is determined to be
\begin{equation}\label{law2o3}
\lambda \sim \gamma^{1/3}H^{2/3},
\end{equation}
where Eq.(\ref{law2o3}) holds
when $H\gg H_c\equiv 12 \sqrt{3\kappa} C^{-3/2} \Gamma/K$.
In Fig.\ref{graph:expfit}., the experimental data of
Allain and Limat \cite{allain} is plotted together with the
theoretical curve given by Eq.(\ref{lambdastar}), where the value
$\gamma=5 \times 10^{-4}$ [m] was used to best fit the experimental data.
Using the experimental data  $(\phi_0-\phi)/\phi_0 \sim 0.1$  in
\cite{allain} and assuming $\alpha$ in Eq.(\ref{cphi}) is of order unity,
we obtain $C_\infty \sim 0.1$.  Thus, from
the fitting value of $\gamma$, $\Gamma/K$ is estimated
as $10^{-8}$ [m], which is of the same order as the length scale of
the microscopic inhomogeneity, i.e., the radius of colloidal particle.
We believe that this result demonstrates the consistency of our theory.



We here comment on the pattern selection of multiple cracks
in heated glass plates recently studied by Yuse et al. \cite{yusesano:multi}.
In their experimental set-up,
many seeds of cracks were prepared in the bottom of a heated glass plate
which is dipped into water. Then, the spacing $\lambda$ of cracks
which finally survive  was measured for the thermal diffusion length $d$.
Motivated by this experiment, Hayakawa carried out numerical simulations
of a spring network model
and found the  scaling relations $\lambda \sim d^{2/3}$ for the limit of
large $d$ \cite{hayakawa:multi}.
This result can easily be interpreted by replacing $H$
in the model for directionally drying fracture with $d$.


Let us turn again to directionally drying fracture.
In the analysis developed above, we postponed two problems.
First, the diffusion length $\xi$ was assumed
to be much larger than the horizontal width $L$.
When $\xi$ is smaller than $L$, the effect of
the evaporation at crack surfaces must be taken into account.
In such a case, the above simple picture cannot be applied to
the crack formation process because inhomogeneity of the water volume
fraction arises. In this case, the crack spacing may have some distribution.
Still, we believe
that the scaling relation given by Eq.(\ref{law2o3}) will give
a first order approximation for the averaged value  of crack spacings.
Second, we assumed that the separation of the strip
from the vertical boundary occurs before the appearance of cracks.
However, even in the case that the separation from the boundaries never occurs,
the vertical boundaries of the slice of material lying between cracks
after their formation will satisfy the stress free boundary conditions.
Therefore, by applying the above discussion to such a slice,
we can again obtain the result Eq.(\ref{law2o3}).


Finally, we briefly discuss the crack patterns which appear subsequent to
the crack formation.  As soon as a crack is formed, it extends along
the $y$ direction and is arrested
 at the position where the energy release rate
for the crack is equal to the surface energy.
After an evaporation time $D/J^2$,
there is no crack formation at the front surface.
Then, one may expect that the individual crack tips
will all extend to the same position.
However, as shown by H-A. Bahr et al.\cite{bahr},
a zig-zag type or more complicated
arrangement occurs when a certain condition is satisfied.
In fact, we can see such patterns in the experiment performed
by Allain and Limat \cite{allain}.
Further, in their experiment, a more interesting crack pattern appeared
due to secondary branching from the elongating cracks.
Such patterns will be studied by devising an extended version of
the present model.

We acknowledge
M. Sano, A. Nakahara and  K. Sekimoto for showing us experiments of
drying fracture and stimulating discussions.
We also thank F.Takagi and Y.Hayakawa for valuable discussions.
The Supercomputer Center, Institute for Solid State Physics,
University of Tokyo is also acknowledged
for allowing us to use FACOM VPP500.
This research was supported in part by JSPS
Research Fellowships for Young Scientists and
by the Japanese Grant-in-aid for Science Research Fund from
the Ministry of Education, Science and Culture
Nos. 07238206 and 07640505.


\begin{figure}
\caption{Schematic view of experimental set-up.
The axes $x,y,z$ are defined as seen.
The drying process takes place only at the front surface,
whose dimension is defined as $L \times H$.
Cracks originate at the surface and extend in the $y$ direction.
}
\label{graph:expset}
\end{figure}

\begin{figure}
\caption{Schematic view of effective one dimensional model for drying fracture
describing fracture pattern formation at the front surface
(see Fig.\protect\ref{graph:expset}).
Each component (circle) is connected with its nearest neighbor components
by $k_1$ springs and with upper and lower plates by $k_2$ springs.
In the drying process, $k_1$ springs may break.
}
\label{graph:effecone}
\end{figure}

\begin{figure}
\caption{
Interval of periodic patterns in drying fracture (vertical axis)
vs. height of strips (horizontal axis).
The solid line ($\gamma^{1/3} H^{2/3}$) is the theoretical curve,
where $\gamma^{1/3}=8.2$ is determined to fit the experimental data.
Filled circles with error bars are experimental data
taken from Allain and Limat \protect\cite{allain}.
The unit length for each axes is $\mu m$.
}
\label{graph:expfit}
\end{figure}


\begin{thebibliography}{99}

\bibitem{herrman}
{\it Statistical Models for the Fracture in Disordered Media},
edited by H.J.Herrmann and S.Roux (North-Holland, Amsterdam, 1990).

\bibitem{griffith}
A.A.Griffith, Phil.Trans.Roy.Soc. {\bf A221}, 163 (1920).

\bibitem{fineberg}
J.Fineberg, S.P.Gross, M.Marder and H.L.Swinny,
Phys.Rev.Lett. {\bf 67}, 457 (1991).

\bibitem{yusesano:vib}
A.Yuse and M.Sano, Nature (London) {\bf 362}, 329 (1993).

\bibitem{allain}
C.Allain and L.Limat, Phys.Rev.Lett. {\bf 74}, 2981 (1995).

\bibitem{nakahara}
A.Nakahara, private communications;
Y.Mitsui and M.Sano, private communications.

\bibitem{crosshohenberg}
M.C.Cross and P.C.Hohenberg,
 Rev.Mod.Phys. {\bf 65}, 851 (1993).

\bibitem{meakin}
P.Meakin, Thin Solid Films {\bf 151}, 165 (1987).

\bibitem{hayakawa:vib}
Y.Hayakawa, Phys.Rev. {\bf E49}, R1804 (1994).

\bibitem{yusesano:multi}
A.Yuse,M.Sano and Y.Couder, private communication.

\bibitem{hayakawa:multi}
Y.Hayakawa, Phys.Rev. {\bf E50}, R1748 (1994).

\bibitem{bahr}
H.-A.Bahr, U.Bahr and A.Petzold, Europhys. Lett. {\bf 19}, 485 (1992).


\end{thebibliography}
\end{document}